\documentclass[12pt]{iopart}
\usepackage{graphicx}% Include figure files
\usepackage{dcolumn}% Align table columns on decimal point
\usepackage{bm}% bold math

\begin{document}

\vspace {-1.5cm} 
\title{High-$p_{T}$ electron distributions in d+Au and p+p collisions at RHIC}
\author{A. A. P. Suaide for the STAR Collaboration\footnote{For the full author list and %%@
acknowledgements see Appendix "Collaborations" in this volume.}}
\address{Instituto de F\'isica, Universidade de S\~ao Paulo \\ Po. Box 66318, 05315-970, %%@
S\~ao Paulo, SP, Brazil}

\date{\today}

\begin{abstract}
We present preliminary  measurements of electron and positron spectra in d+Au and p+p %%@
collisions  at $\sqrt{s_{NN}}=200$ GeV for $1.5 < p_{T} < 7.0$ GeV/c. These measurements were %%@
carried  out using the STAR Time Projection Chamber (TPC) and the Barrel Electromagnetic  %%@
calorimeter (EMC).   Overall hadron  rejection factors in the range of $10^{5}$ have been %%@
achieved.  In this work we describe the measurement technique used to discriminate electrons  %%@
from hadrons and compare the results for single electron spectra with Pythia based pQCD %%@
calculations for electrons from heavy-quark semi-leptonic decays.
\end{abstract}

The primary electron spectrum over a sufficiently broad $p_{T}$ range provides a measurement  %%@
of charm and beauty production at RHIC energies. In heavy ion collisions, these heavy  quark %%@
production rates are expected to be an important diagnostic of the dense system formed in the %%@
collision. In particular,  comparative measurements in p+p, d+Au and Au+Au will provide %%@
important sensitivity to  the initial state gluon densities in these systems \cite{gluon} and %%@
medium effects  such as heavy quark energy loss. The suppression of small angle gluon %%@
radiation for heavy quarks would decrease the amount of energy loss (dead cone effect) %%@
\cite{dead} and, if gluon bremsstrahlung is indeed the main mechanism of quark energy loss, %%@
the suppression of heavy quark mesons at high-$p_{T}$ is expected to be smaller than that one %%@
observed for charged hadrons at RHIC \cite{highpt}. This comparison is an important check of %%@
the quenching mechanism at heavy-ion collisions. Moreover, measuring open charm and beauty %%@
production at RHIC provides essential reference data  for studies of color screening via %%@
quarkonium suppression \cite{jpsi}. 

The results  presented in this work were obtained with the STAR detector using the Time %%@
Projection Chamber (TPC) and the first EMC patch installed for the 2003 RHIC run, which %%@
consisted of 60 modules, half of the full planned detector, with coverage from  $0< \eta < 1$ %%@
and $\Delta \phi = 360^{o}$. Each one of the EMC modules is divided into 40 towers with %%@
spatial coverage of $(\Delta\eta, \Delta\phi) = (0.05, 0.05)$. The tower depth is 21  %%@
radiation lengths ($X_{0}$).  A Shower Max Detector (SMD) is located approximately 5 $X_{0}$ %%@
deep inside the calorimeter module and allows to measure the electromagnetic shower shape and %%@
position with high precisison ($\Delta\eta,\Delta\phi)\sim(0.007,0.007)$. Details about the %%@
detectors used in this analysis can be found in Ref. \cite{star}. 

The process of electron identification using the STAR barrel calorimeter is based  on a %%@
pre-selection of electron candidates from the TPC $dE/dx$ measurement. Electrons in the %%@
momentum range between 1.5 and 8 GeV/c have slightly higher $dE/dx$ values when compared to %%@
hadrons. A $dE/dx$ cut in this momentum range provides initial discimination  power on the %%@
order of $e/h \sim 500$ with high efficiency.

\begin{figure} [hc]
  \vspace{-0.5 cm}
  \begin{center}
    \includegraphics[width=0.75\textwidth]{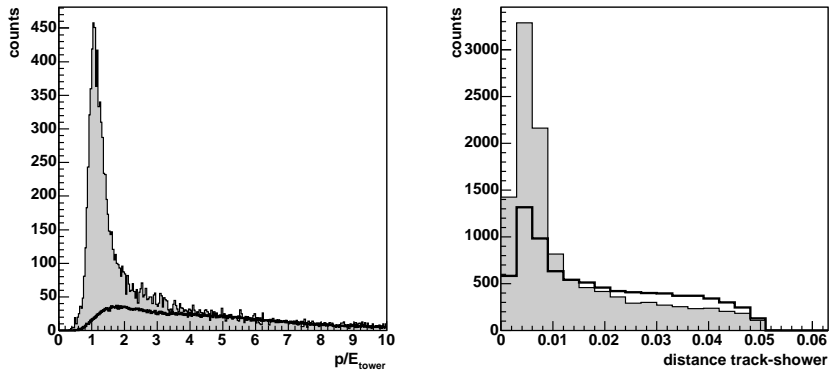}
  \end{center}
  \vspace{-0.5 cm}
  \caption{Left: $p/E_{tower}$ distributions. Right: Distance between extrapolated 
           track and Shower Max detector shower position. Shadowed histograms are the
           distributions for electrons and the non-shadowed ones are distributions for %%@
hadrons.
           \label{figPoverE}}
\end{figure}

After the electron candidates are selected, they are extrapolated to the EMC detector  and %%@
the energy deposited in the tower hit by the candidate is compared to its momentum.  %%@
Electrons should show a peak at $p/E_{tower}\sim 1$. Hadrons have a wider distribution of  %%@
$p/E_{tower}$. Figure \ref{figPoverE}-left shows the $p/E_{tower}$ spectrum for the electron %%@
candidates  in which it is possible to see a well defined electron peak. The residual %%@
hadronic background  is shown as a solid line in the spectrum. After hadronic background %%@
subtraction the electron  peak is not centered at 1 due to the energy leakage to neighboring %%@
towers. The amount of leakage  depends on the distance to the center of the tower hit by the %%@
electron and it is well descibed by GEANT simulations of the detector response.

The shower max detector plays an important role in the electron identification procedure.  In %%@
general, hadronic showers are not well developed compared to electromagnetic showers in the  %%@
shower max region of the EMC. The resulting differences are used to enhance the  electron %%@
discrimination power. The procedure used in this analysis was to set high  thresholds in the %%@
shower max shower reconstruction. Electrons will have showers reconstructed   well with these %%@
cuts while hadrons will have very low efficiency shower reconstruction. We also compare  the %%@
distance of the extrapolated particle to the reconstructed shower. Because of the poorly  %%@
developed showers in the case of hadrons, this distance will have a much wider distribution, %%@
as seen in Figure \ref{figPoverE}-right. The overall electron identification efficiency was %%@
obtained by embedding simulated electrons into real events and was found to be $\sim50$\% and %%@
$p_{T}$ independent for electrons with $p_{T}>2$ GeV/c.

The electrons measured originate from various different sources. We classify two categories: %%@
(i) the physics signal, composed from semileptonic heavy quark decays and Drell-Yan processes %%@
and (ii) the background sources. The background electrons are mainly from secondary electrons %%@
(photon conversions and Dalitz decays of light vector mesons) and hadrons misidentified as %%@
electrons. This  background should be removed from the spectra in order to address the %%@
physics signal.

\begin{figure}[h]
  \vspace{-0.5 cm}
  \begin{center}
    \includegraphics[width=0.75\textwidth]{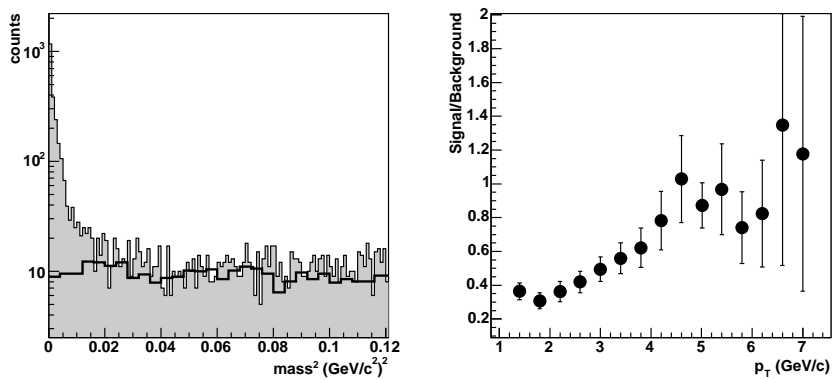}
  \end{center}
  \vspace{-0.5 cm}
  \caption{Left: $mass^{2}$ spectra for e$^{+}$e$^{-}$ pairs (histogram) and same
  charge sign pair (line). Right: Signal to background ratio for electron as a
  function of electron $p_{T}$.\label{figConv}}
\end{figure}

Most of photon conversion and $\pi^{0}$ Dalitz decays can be  removed by calculating the %%@
invariant mass spectrum of di-electrons. Figure \ref{figConv}-left shows the  $m^{2}$ %%@
spectrum for opposite and same charge electron pairs. A cut of $m^{2}< 0.02$ %%@
(GeV/c$^{2}$)$^{2}$  removes most of the photon conversion and Dalitz decay electrons. The %%@
remaining background, mainly composed of $\eta$, $\omega$, $\phi$ and $\rho$ decays, was %%@
estimated from  Pythia \cite{pythia} and HIJING \cite{hijing} simulations and it is on the %%@
level of a few percent of the total background.  Figure \ref{figConv}-right shows the ratio %%@
between the physics signal to background electrons. The overall signal to background ratio %%@
improves substantially at high-$p_{T}$.

The hadron contamination was estimated by selecting hadrons using TPC dE/dx and computing  %%@
how many of them are identified as electrons in the EMC. Residual hadronic contamination is %%@
on the order of 3\% for $p_{T} = 2$ GeV/c and 8\% for $p_{T} = 6$ GeV/c. By combining the TPC %%@
and EMC it is possible to achieve an $e/h$ discrimination power on the order of  $10^{5}$ %%@
while maintaining an electron identification efficiency around 50\%.

\begin{figure} [h]
  \vspace{-0.5 cm}
  \begin{center}
   \resizebox{\textwidth}{!}{
     \includegraphics[width=0.5\textwidth]{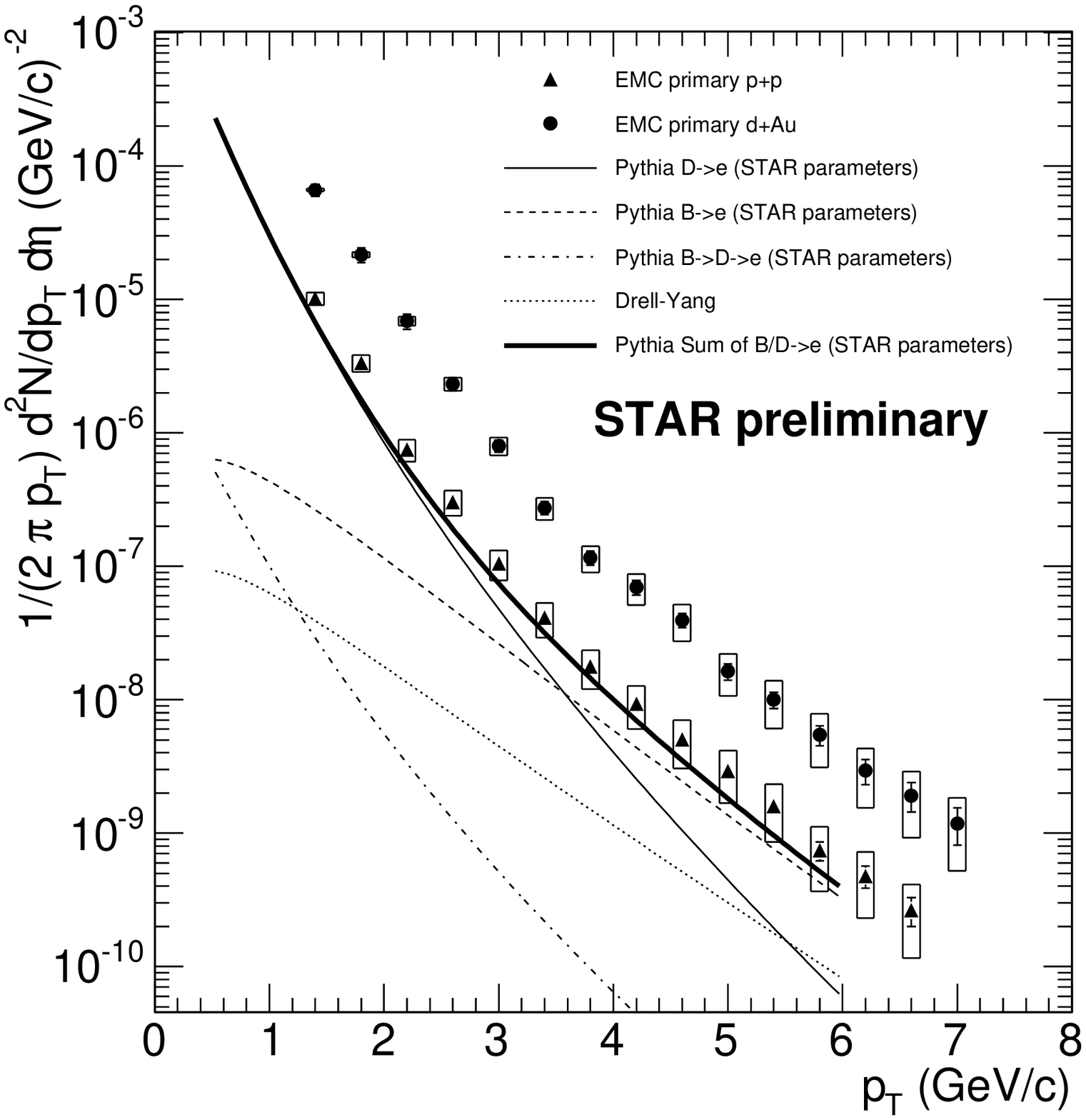}
      \includegraphics[width=0.5\textwidth]{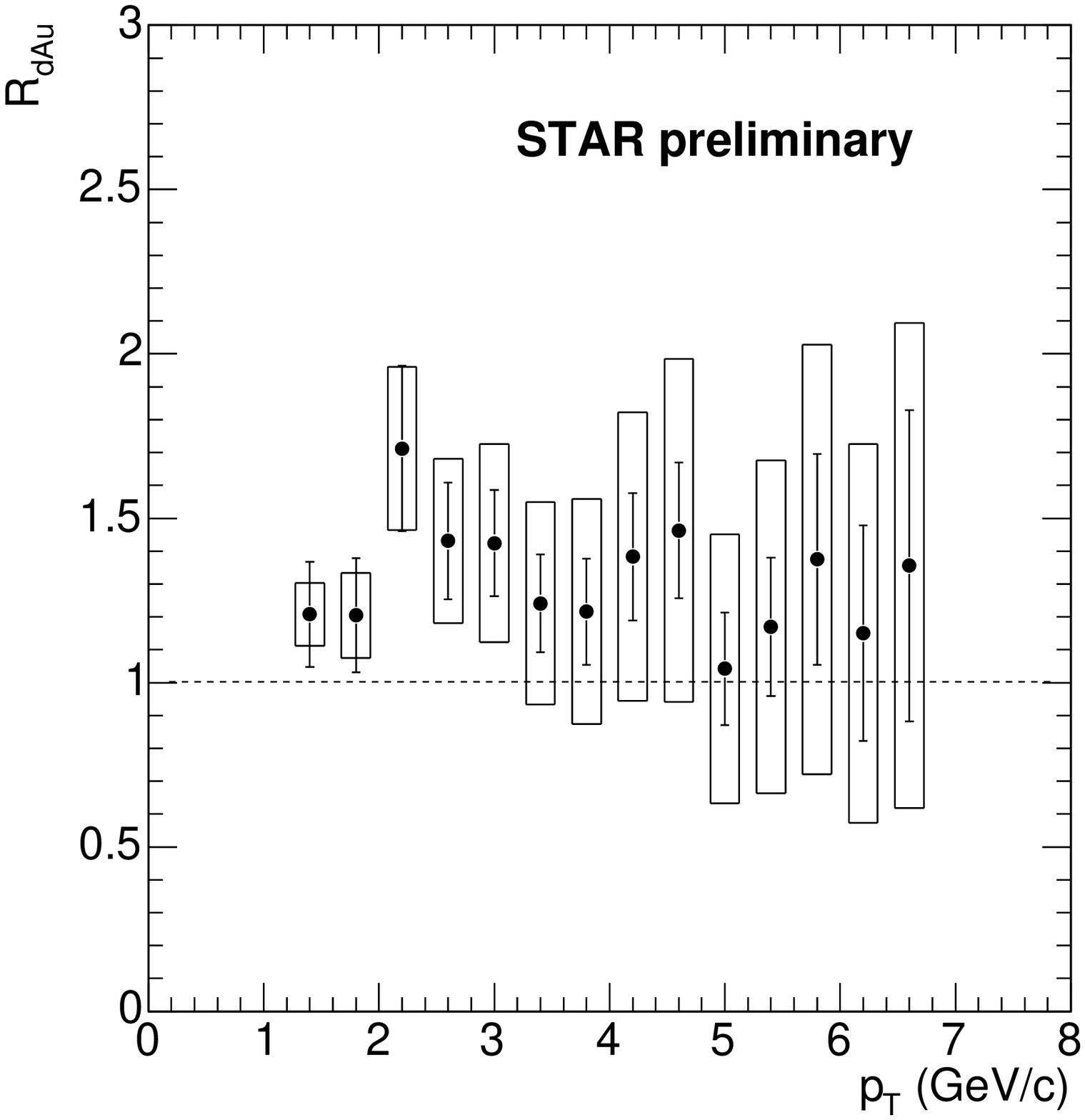}}
 \end{center}
  \vspace{-0.5 cm}
\caption{Left: Background subtracted electron spectra for d+Au (circles) and p+p (triangles) %%@
collisions. The error bars indicate the statistical errors and the boxes show the systematic %%@
uncertainties. The lines show Pythia simulations (see text for parameters). Right: $R_{dAu}$ %%@
for electrons at $\sqrt{s_{NN}}=200$ GeV. There is an overall normalization error of 17.4\% %%@
on the unity that is not shown in the figure.\label{primary}}
\end{figure}

Figure \ref{primary}-left shows the primary electron spectra for d+Au and p+p collisions at  %%@
$\sqrt{s_{NN}}=200$ GeV. The error bars are statistical errors and the boxes depict the %%@
systematic uncertainties.

The lines in the Figure \ref{primary}-left show the electron spectra prediction for p+p %%@
collisions  from Pythia simulations. The thick solid line is the total electron yield %%@
prediction while  the thin solid and dashed lines are predictions for electrons from D  and B %%@
mesons decays  respectively. The dash-dotted line is the contribution to the electron %%@
spectrum for B mesons  decaying into D mesons before decaying to electrons and the %%@
contribution to the total  yield is negligible. The dotted line is the contribution from %%@
Drell-Yan to the electron yield. The Pythia  parameters used in the current simulations are: %%@
$\langle K_{T}\rangle = 2$ GeV/c, $m_{C}=1.7$ GeV/c$^2$, $K=2.2$, CTEQ5M1 and $PARP(67)=4$. %%@
It is important to notice that the Pythia simulation is not a fit to the data but just a %%@
representation of what may be the sources  of electrons observed and the parameters used are %%@
still under investigation. We note, however, that electrons at moderate to high $p_{T}$  %%@
($p_{T}>3.5$ GeV/c) have a significant to dominant contribution from B decays, being the %%@
first RHIC measurement sensitive to beauty cross section. Figure \ref{primary}-right shows %%@
the ratio, $R_{dAu}$, of the d+Au and p+p spectra, normalized for the number of binary %%@
collisions, as a function of $p_{T}$.  It is important to notice that the electron $R_{dAu}$ %%@
at a given $p_{T}$ arises from a wide heavy-flavor $p_{T}$ range. The ratio is approximately %%@
consistent with unity for the entire momentum range suggesting that the electron production %%@
in d+Au collisions follow a simple binary scaling from p+p collisions. A small Cronin type %%@
enhancement can not be ruled out. The magnitude of the Cronin effect for heavy quark mesons %%@
is not significantly different from that of light quark hadrons \cite{dAu}.

\end{document}